\documentclass[11pt]{article} 

\usepackage[utf8]{inputenc} 

\usepackage{geometry} 
\usepackage{graphicx} 

\usepackage{booktabs} 
\usepackage{array} 
\usepackage{paralist} 
\usepackage{verbatim} 
\usepackage{subfig} 

\usepackage{amssymb}
\usepackage{amsmath}
\usepackage{amsthm}
\usepackage{mathrsfs}

\theoremstyle{plain}
\newtheorem{thm}{Theorem}[section]

\newtheorem{exm}[thm]{Example}
\newtheorem{defn}[thm]{Definition}
\newtheorem{rem}[thm]{Remark}

\usepackage{fancyhdr} 
\usepackage{sectsty}
\allsectionsfont{\sffamily\mdseries\upshape} 

\usepackage[nottoc,notlof,notlot]{tocbibind} 
\usepackage[titles,subfigure]{tocloft} 

\title{Self-adjointness and conservation laws of difference equations}
\author{Linyu Peng}
\date{} 

\begin{document}
\maketitle

\abstract{A general theorem on conservation laws for arbitrary difference equations is proved. The theorem is based on an introduction of an adjoint system related with a given difference system, and it does not require the existence of a difference Lagrangian. It is proved that the system, combined by the original system and its adjoint system, is governed by a variational principle, which inherits all symmetries of the original system. Noether's theorem can then be applied. With some special techniques, e.g. self-adjointness properties, this allows us to obtain conservation laws for difference equations, which are not necessary governed by Lagrangian formalisms.}

\section{Introduction}
Symmetries of variational principles are naturally symmetries of the associated Euler-Lagrange equations. A connection between such type of symmetries and conservation laws for differential equations is estabilished via Noether's theorem \cite{Bluman1989,Ibragimov1985,Noether1918,Olver1993}. However, Noether's theorem has difficulty in applications to arbitrary differential equations. This is overcome by Ibragimov \cite{Ibragimov2007} by defining an adjoint system for arbitrary differential equations, and constructing a Lagrangian for a given differential system together with its adjoint system. Symmetries of a given differential system can then be extended to variational symmetries for the Lagrangian. At this stage, Noether's theorem is hence appliable.

Geometric methods, especially symmetry analysis, as are used for investigating differential equations, have been applied to difference equations over the last decades, see \cite{Hydon2014,Levi2006,Yamilov2006}. A discrete version of Noether's theorem exists, see for example \cite{Dorod2001,Grant2013,Maeda1981,Peng2014}. In the present paper, we attempt to generalize Ibragimov's method for constructing conservation laws from differential systems to difference systems. For a system of difference equations, its transformation groups of symmetries can be extended to groups of variational symmetries for a Lagrangian governing the original system itself together with its adjoint system. Thus, Noether's theorem can be used to construct conservation laws for the combined system. For (strict, quasi, weak) self-adjoint systems, it is possible to transfer such conservation laws to conservation laws of the original system. This procedure can also be realised once special solutions of the adjoint system are known.

\section{Ibragimov's conservation laws for differential equations}
Let $x=(x^1,x^2,\ldots,x^p)$ and $u=(u^1,u^2,\ldots,u^q)$ be $p$ independent variables and $q$ dependent variables, respectively. Let $J=(j_1,j_2,\ldots,j_p)$, and let $u^{\alpha}_J$ denote $|J|^{\operatorname{th}}$-order partial derivatives of $u$. Here $|J|=j_1+j_2\cdots+j_p$ and
\begin{equation}
u_J^{\alpha}:=\frac{\partial ^{|J|}u^{\alpha}}{(\partial x^1)^{j_1}(\partial x^2)^{j_2}\cdots(\partial x^p)^{j_p}}.
\end{equation}
Consider a linear differential operator (the Lie-B\"{a}cklund operator),
\begin{equation}
\begin{aligned}
X&=\xi^iD_i+\left(\eta^{\alpha}-\xi^iu_i^{\alpha}\right)\frac{\partial}{\partial u^{\alpha}}+\cdots+\sum_{\alpha,J}D_J\left(\eta^{\alpha}-\xi^iu_i^{\alpha}\right)\frac{\partial}{\partial u_J^{\alpha}}+\cdots\\
&=\xi^i\frac{\partial}{\partial x^i}+\eta^{\alpha}\frac{\partial}{\partial u^{\alpha}}+\cdots.
\end{aligned}
\end{equation}
Here $\xi^i=\xi^i(x,[u])$ and $\eta^{\alpha}=\eta^{\alpha}(x,[u])$ are smooth functions, where $[u]$ denotes $u$ and its derivatives. The operator $D_i$ is the total derivative with respect to $x^i$ and $D_J$ is a composite of total derivatives. The set of all such differential operators is a Lie algebra equipped with the usual Lie bracket between two vector fields.

For a system of partial differential equations
\begin{equation}\label{de0}
F_{\alpha}(x,[u])=0,~\alpha=1,2,\ldots,q,
\end{equation}
the adjoint system is given by
\begin{equation}\label{ade0}
0=F_{\alpha}^{*}(x,[u],[v]):=\bold{E}_{u^{\alpha}}\left(v^{\beta}F_{\beta}\right).
\end{equation}
The Euler operator $\bold{E}_{u^{\alpha}}$ is defined as
\begin{equation}
\bold{E}_{u^{\alpha}}:=\frac{\partial}{\partial u^{\alpha}}-D_i\frac{\partial}{\partial u^{\alpha}_i}\cdots+(-D)_{J}\frac{\partial}{\partial u_J^{\alpha}}+\cdots
\end{equation}
where $(-D)_{J}=(-1)^{|J|}D_J$. The system of differential equations (\ref{de0}) and (\ref{ade0}) corresponds to a Lagrangian $L(x,[u],[v])=v^{\alpha}F_{\alpha}(x,[u])$, and we have
\begin{equation}
\begin{aligned}
&\bold{E}_{u^{\alpha}}(L)=F^*_{\alpha}(x,[u],[v]),\\
&\bold{E}_{v^{\alpha}}(L)=F_{\alpha}(x,[u]).
\end{aligned}
\end{equation}
A differential system is said to be self-adjoint if the system $F_{\alpha}^{*}(x,[u],[u])=0$ is identical to the original one.

If a differential operator $X=\xi\frac{\partial}{\partial x^i}+\eta^{\alpha}\frac{\partial}{\partial u^{\alpha}}$ is a symmetry generator for the equations (\ref{de0}), there always exists a variational symmetry generator for the Lagrangian $L$, namely $Y=\xi\frac{\partial}{\partial x^i}+\eta^{\alpha}\frac{\partial}{\partial u^{\alpha}}+\eta^{\alpha}_*\frac{\partial}{\partial v^{\alpha}}$. The coefficient $\eta_*^{\alpha}$ is determined by the generator $X$. Hence, by using Noether's theorem and the generator $Y$, we can construct conservation laws for the combination of (\ref{de0}) and (\ref{ade0}). With knowledge of particular solutions of $v$ or self-adjointness of the original system, we can get conservation laws of the original equations (\ref{de0}). For more details, consult \cite{Ibragimov2007}.

\begin{exm}[\cite{Ibragimov2007}]
Consider the Korteweg-de Vries equation
\begin{equation}
u_t=u_{xxx}+uu_x
\end{equation}
and its adjoint equation
\begin{equation}
v_t=v_{xxx}+uv_x.
\end{equation}
It is easy to see that the KdV equation is self-adjoint.
These two equations are governed by the following Lagrangian
\begin{equation}
L=v\left(u_t-uu_x-u_{xxx}\right).
\end{equation}
Symmetries of a scaling transformation with generator
\begin{equation}
X=-3t\frac{\partial}{\partial t}-x\frac{\partial}{\partial x}+2u\frac{\partial}{\partial u}
\end{equation}
will be extended to variational symmetries of the Lagrangian $L$, that is,
\begin{equation}
Y=-3t\frac{\partial}{\partial t}-x\frac{\partial}{\partial x}+2u\frac{\partial}{\partial u}-v\frac{\partial}{\partial v}.
\end{equation}
Conservation law $D_tP^1+D_xP^2$ obtained through Noether's theorem is hence given by
\begin{equation}
P^1=(3tu_{xxx}+3tuu_x+xu_x+2u)v
\end{equation}
and
\begin{equation}
\begin{aligned}
P^2=&-(2u^2+xu_t+3tuu_t+4u_{xx}+3tu_{txx})v\\
&+(3u_x+3tu_{tx}+xu_{xx})v_x-(2u+3tu_t+xu_x)v_{xx}.
\end{aligned}
\end{equation}
Setting $v=u$ and transfering the terms of the form $D_x(\ldots)$ from $P^1$ to $P^2$, we get
\begin{equation}
P^1=u^2,~P^2=u_x^2-2uu_{xx}-\frac{2}{3}u^3.
\end{equation}
\end{exm}

\section{Self-adjointness and conservation laws of difference equations}
Let us consider, in general, a given difference system with independent variables
$n=(n^1,n^2,\ldots,n^p)\in \mathbb{Z}^p$, and dependent variables
$u=(u^1,u^2,\ldots,u^q)\in U\subset \mathbb{R}^q$. Its solutions
$u=f(n)$ can be viewed as lying on sections $s(n)=(n,f(n))$ of
the trivial bundle $\pi:\mathbb{Z}^p\times U\rightarrow\mathbb{Z}^p$
with $\pi(n,u)=n$, implying that $\mathbb{Z}^p$ is viewed as the
base space.
The shift operator (or map) $S$ is defined as
\begin{equation}
\begin{aligned}
S_k:n^i\mapsto n^i+\delta_k^i,~~k=1,2,\ldots,p,
\end{aligned}
\end{equation}
with $\delta^i_k$ the Kronecker delta. Let $1_k$ be the $p$-tuple
with only one nonzero entry, which is $1$, at the $k^{\operatorname{th}}$ place. Then
the $k^{\operatorname{th}}$-shift operator and the naturally extended shift operator
to a function $f(n)$ are respectively given by
\begin{equation}
\begin{aligned}
S_k:n\mapsto n+1_k
\end{aligned}
\end{equation}
and
\begin{equation}
\begin{aligned}
S_k:f(n)\mapsto f(n+1_k).
\end{aligned}
\end{equation}
We sometimes use the notation $S_{1_k}$ instead of $S_k$, and the
composite of shifts using multi-index notation is given by
$S_J=S_1^{j_1}S_2^{j_2}\cdots S_p^{j_p}$, where
$J=(j_1,j_2,\ldots,j_p)$ is a $p$-tuple. Moreover, we can define the
inverse of the shift map $S_{1_k}$ as $S_{-1_k}:n\to n-1_k$. The
inverse map $S_{-J}$ of the composite of shifts is similarly
defined.
We construct the prolongation bundle, $\bold{pr}^{(\infty)}(\mathbb{Z}^p\times U)$,
 which has induced local coordinates \cite{Peng2013,Peng2014}
 \begin{equation}
 (n,u,u_{\{1\}},u_{\{2\}},\ldots,u_{\{-1\}},u_{\{-2\}},\ldots),
 \end{equation}
where $u_{\{k\}}$ denotes all $k^{\operatorname{th}}$-order
shifts of $u$. For example, we have
$u_{\{1\}}=\{u^{\alpha}_{1_i}\}$, where
$u^{\alpha}_{1_i}=S_if^{\alpha}(n)$. 

Consider a system of difference equations
\begin{equation}\label{de}
\mathcal{A}=\{F_{\alpha}(n,[u])=0,~\alpha=1,2,\ldots,q\},
\end{equation}
where $F_{\alpha}$ are analytic functions with respect to $[u]$, which denotes $u$ and its shifts. A one-parameter group $G$ of transformations related to this system is a symmetry group, if and only if the associated infinitesimal generator $X$ satisfies
\begin{equation}\label{sccc}
\bold{pr}^{(\infty)}X(F_{\alpha})=0 \text{ on solutions of (\ref{de})}.
\end{equation}
Often we write an infinitesimal generator as
\begin{equation}
X=Q^{\alpha}(n,[u])\frac{\partial}{\partial u^{\alpha}},
\end{equation}
where $Q^{\alpha}(n,[u])$ are called characteristics with respect to a group of symmetries.
Its prolongation is 
\begin{equation}
\bold{pr}^{(\infty)}X=X+\cdots+\sum_{\alpha,J}S_JQ^{\alpha}\frac{\partial}{\partial u^{\alpha}_J}+\cdots.
\end{equation}

\begin{rem}
Since $F_{\alpha}$ are analytic functions, the symmetry criterion (\ref{sccc}) can be re-stated as that a vector field $X$ generates a group of symmetries for a difference system $\mathcal{A}$ if and only if
there exist difference operators $\mathscr{B}_{\alpha\beta}=B^J_{\alpha\beta}(x,[u])S_J$ such that 
\begin{equation}
\bold{pr}^{(\infty)}X(F_{\alpha})=\sum_{\beta}\mathscr{B}_{\alpha\beta}\left(F_{\beta}\right).
\end{equation}
\end{rem}

\begin{thm}
Assume that $X_1=Q_1^{\alpha}\frac{\partial}{\partial u^{\alpha}}$ and $X_2=Q_2^{\alpha}\frac{\partial}{\partial u^{\alpha}}$ are two infinitesimal generators of symmetries for (\ref{de}), then so does $[X_1,X_2]$.
\end{thm}
\begin{proof}
First we prove that $\bold{pr}^{(\infty)}[X_1,X_2]=[\bold{pr}^{(\infty)}X_1,\bold{pr}^{(\infty)}X_2]$. Its right-hand side is
\begin{equation*}
[\bold{pr}^{(\infty)}X_1,\bold{pr}^{(\infty)}X_2]=\sum_{J_1,J_2}\left((S_{J_1}Q_1^{\alpha})\frac{\partial (S_{J_2}Q_2^{\beta})}{\partial u_{J_1}^{\alpha}}-(S_{J_1}Q_2^{\alpha})\frac{\partial (S_{J_2}Q_1^{\beta})}{\partial u_{J_1}^{\alpha}}\right)\frac{\partial}{\partial u_{J_2}^{\beta}},
\end{equation*}
while $[X_1,X_2]$ is given by
\begin{equation*}
[X_1,X_2]=\sum_{J_1}\left((S_{J_1}Q_1^{\alpha})\frac{\partial Q_2^{\beta}}{\partial u_{J_1}^{\alpha}}-(S_{J_1}Q_2^{\alpha})\frac{\partial Q_1^{\beta}}{\partial u_{J_1}^{\alpha}}\right)\frac{\partial}{\partial u^{\beta}}.
\end{equation*}
It is not difficult to find out that for any $J_2$, the following calculation is valid
\begin{equation*}
\begin{aligned}
S_{J_2}\sum_{J_1}&\left((S_{J_1}Q_1^{\alpha})\frac{\partial Q_2^{\beta}}{\partial u_{J_1}^{\alpha}}-(S_{J_1}Q_2^{\alpha})\frac{\partial Q_1^{\beta}}{\partial u_{J_1}^{\alpha}}\right)\\
&=\sum_{J_1}\left((S_{J_1+J_2}Q_1^{\alpha})\frac{\partial (S_{J_2}Q_2^{\beta})}{\partial u_{J_1+J_2}^{\alpha}}-(S_{J_1+J_2}Q_2^{\alpha})\frac{\partial (S_{J_2}Q_1^{\beta})}{\partial u_{J_1+J_2}^{\alpha}}\right)\\
&=\sum_{J_1}\left((S_{J_1}Q_1^{\alpha})\frac{\partial (S_{J_2}Q_2^{\beta})}{\partial u_{J_1}^{\alpha}}-(S_{J_1}Q_2^{\alpha})\frac{\partial (S_{J_2}Q_1^{\beta})}{\partial u_{J_1}^{\alpha}}\right),
\end{aligned}
\end{equation*}
that is, $\bold{pr}^{(\infty)}[X_1,X_2]=[\bold{pr}^{(\infty)}X_1,\bold{pr}^{(\infty)}X_2]$. Therefore, the symmetry criterion implies that 
\begin{equation*}
\begin{aligned}
\bold{pr}^{(\infty)}[X_1,X_2](F_{\alpha})=[\bold{pr}^{(\infty)}X_1,\bold{pr}^{(\infty)}X_2](F_{\alpha}),
\end{aligned}
\end{equation*}
whose right-hand side can be written as the form $\sum_{\beta,J}B_{\alpha\beta}^JS_J(F_{\beta})$. This finishes the proof. 
\end{proof}
The following equality is used during the proof that, for any $J$, 
\begin{equation}
\begin{aligned}
\bold{pr}^{(\infty)}X(S_J(F_{\alpha}))&=\sum_{\beta,I}Q_{I}^{\beta}\frac{\partial (S_JF_{\alpha})}{\partial u_{I}^{\beta}}\\
&=
\sum_{\beta,I}S_J\left(Q_{I-J}^{\beta}\frac{\partial F_{\alpha}}{\partial u_{I-J}^{\beta}}\right)\\
&=S_J\left(\bold{pr}^{(\infty)}X(F_{\alpha})\right).
\end{aligned}
\end{equation}

A conservation law is defined as the vanishment of a difference divergence expression 
\begin{equation}
\operatorname{Div}^{\vartriangle}P:=\sum_{i=1}^p (S_i-\operatorname{id})P^i
\end{equation}
on solutions of $\mathcal{A}$. 

For a difference variational problem 
\begin{equation}
\mathscr{L}[u]=\sum_nL(n,[u]),
\end{equation}
the invariance criterion reads infinitesimally as
\begin{equation}
\bold{pr}^{(\infty)}X(L_n)=\operatorname{Div}^{\vartriangle}R,
\end{equation}
for some $p$-tuple $R$. Here we write $L_n=L(n,[u])$. The associated difference Euler-Lagrange equations $\bold{E}^{\vartriangle}_{u^{\alpha}}(L_n)=0$ are obatained by using the difference Euler 
operator
\begin{equation}
\bold{E}^{\vartriangle}_{u^{\alpha}}:=\sum_JS_{-J}\frac{\partial}{\partial u^{\alpha}_J}.
\end{equation}
The invariance of $\mathscr{L}[u]$ implies the invariance of the difference Euler-Lagrange equations \cite{Peng2014}.
 A discrete version of Noether's theorem exists, which establishs a connection between variational symmetries andconservation laws of the difference Euler-Lagrange equations. It has been proved that, for any finite tuple $J$, two equations $\bold{E}^{\vartriangle}_{u^{\alpha}}(L_n)=0$ and $\bold{E}^{\vartriangle}_{u^{\alpha}}(S_JL_n)=0$ are equivalent to each other \cite{Peng2014a}. For Lagrangians being independent from backward shifts, a general form of conservation laws obtained from Noether's theorem is $\operatorname{Div}^{\vartriangle}P=0$ with \cite{Peng2014}
\begin{equation}\label{ncl}
P^i=\sum_{\alpha, J\geq 1_i}Q^{\alpha}_{J-{1_i}}S_{-1_i}\left(\bold{E}^{\vartriangle}_{u^{\alpha}_J}(L_n)\right)-R^i.
\end{equation}
Here the operator $\bold{E}^{\vartriangle}_{u^{\alpha}_J}$ is given by
\begin{equation}
\bold{E}^{\vartriangle}_{u^{\alpha}_J}:=\sum_{I\geq 0}S_{-I}\frac{\partial}{\partial u^{\alpha}_{I+J}}.
\end{equation}

Let $\mathcal{H}$ be a linear operator, that is, it can be written as a polynomial of shift operators whose coefficients are functions on the prolongation bundle. Its adjoint operator $\mathcal{H}^*$ is defined by the following equality
\begin{equation}\label{lao}
v\mathcal{H}[u]=u\mathcal{H}^*[v]+\operatorname{Div}^{\vartriangle}P.
\end{equation}
The equation $\mathcal{H}^*[v]=0$ is called the adjoint equation of $\mathcal{H}[u]=0$. If for any $u(n)$, $\mathcal{H}^*[u]=\mathcal{H}[u]$ holds, the operator $\mathcal{H}$ is said to be (strictly) self-adjoint. Nevertheless, for many cases, $\mathcal{H}^*[u]=0$ and $\mathcal{H}[u]=0$ are equivalent to rather than equal to each other.
\begin{exm}\label{ex32}
The adjoint equation of $u_{n+2}-u_n=0$ is $v_{n-2}-v_n=0$. These two equations are equivalent to each other. If we rewrite the original equation as $u_{n+1}-u_{n-1}=0$, the adjoint equation is then $v_{n-1}-v_{n+1}=0$.
\end{exm}

\begin{defn}
Consider a system of difference equations (\ref{de}), and assume the functions $F_{\alpha}(n,[u])$ are independent from backward shifts of $u$. We define the system of its {\bf adjoint equations} as
\begin{equation}\label{ade}
0=F_{\alpha}^*(n,[u],[v]):=\bold{E}^{\vartriangle}_{u^{\alpha}}\left(v_n^{\beta}F_{\beta}\right).
\end{equation} 
\end{defn}

In the case of linear equations, this definition is equivalent to the one in (\ref{lao}). When $F_{\alpha}$ are linear equations, the adjoint equations are linear with respect to $[v]$ and independent from $[u]$. Otherwise, the adjoint equations are linear with respect to $[v]$, but can be nonlinear in the coupled variables $[u]$ and $[v]$. For linear difference systems, the order of the adjoint system are the same as that of the original system. Nevertheless, in the nonlinear case, the order of the adjoint system is usually higher than the original system.

\begin{defn}
A system of difference equations (\ref{de}) is said to be {\bf self-adjoint} if its adjoint system by the substitution $v=u$,
\begin{equation}
F_{\alpha}^*(n,[u],[u])=0
\end{equation}
holds for all solutions of the original system (\ref{de}).
\end{defn}

\begin{exm}
In Example \ref{ex32}, it is easy to show that the equation $u_{n+2}-u_n=0$ is self-adjoint, since by the substitution $v=u$, the adjoint equation becomes $u_{n-2}-u_n=0$, which is equivalent to the original one.
\end{exm}

\begin{exm}
In general, let us consider the following second-order linear difference equation
\begin{equation}\label{o2de}
a_1(n)u_{n+2}+a_2(n)u_{n+1}+a_3(n)u_n=0.
\end{equation}
Here $a_1(n)a_3(n)\neq0$ for all $n$. Its adjoint equation is
\begin{equation}
 a_3(n)v_n+a_2(n-1)v_{n-1}+a_1(n-2)v_{n-2}=0.
\end{equation}
If the following equalities hold 
\begin{equation}
a_3(n+2)=a_1(n),~a_2(n+1)=a_2(n),~a_1(n)=a_3(n),
\end{equation}
the difference equation (\ref{o2de}) is self-adjoint. Namely, a sufficient (but not necessary) condition for the equation (\ref{o2de}) to be self-adjoint is that 
\begin{equation}\label{o2dec}
a_1(n)=a_3(n)=C_1\frac{1-(-1)^n}{2}+C_2\frac{1+(-1)^n}{2},~a_2(n)=C_3,
\end{equation}
where $C_i$ are constants and $C_1C_2\neq0$.
\end{exm}

In the differential case, the generalizations of self-adjointness to quasi and weak self-adjointnesses are aslo very useful in finding conservation laws \cite{Gandarias2011,Ibragimov2007a}. Weak self-adjointness is also named nonlinear adjointness in \cite{Ibragimov2011}. We adjust such ideas to difference equations.
\begin{defn}
A difference system (\ref{de}) is said to be {\bf quasi self-adjoint} if by a nontrivial substitution $v^{\alpha}=f^{\alpha}([u])$,  the adjoint system 
\begin{equation}
F_{\alpha}^*(n,[u],[f([u])])=0
\end{equation}
holds for all solutions $u$ of the original system.
The original system is said to be {\bf weak self-adjoint} if by a nontrivial substitution $v^{\alpha}=f^{\alpha}(n,[u])$, the adjoint system is satisfied for all solutions $u$ of the original system. Here by nontrivial we mean that not all the functions $f^{\alpha}$ vanish simultaneously. 
\end{defn}

\begin{exm}
Consider the second-order difference equation (\ref{o2de}) again. If we change the condition (\ref{o2dec}) to 
\begin{equation}
a_1(n)=a_3(n)=C_1\frac{1-(-1)^n}{2}+C_2\frac{1+(-1)^n}{2},~a_2(n)=(-1)^nC_3,
\end{equation}
then the difference equation is weak self-adjoint that can be verified by a substitution $v_n=(-1)^nu_n$.
\end{exm}

\begin{thm}
Any difference system (\ref{de}) together with its adjoint system (\ref{ade}) are governed by a Lagrangian.
\end{thm}
\begin{proof}
Let us define a Lagrangian 
\begin{equation}
L_n(n,[u],[v])=v_n^{\alpha}F_{\alpha}(n,[u]).
\end{equation}
Direct calculation shows that
\begin{equation}
\begin{aligned}
&\bold{E}^{\vartriangle}_{v^{\alpha}}(L_n)=F_{\alpha}(n,[u]),\\
&\bold{E}^{\vartriangle}_{u^{\alpha}}(L_n)=F^*_{\alpha}(x,[u],[v]).
\end{aligned}
\end{equation}
Thus, the difference Euler-Lagrange equations cover the original system (\ref{de}) and its adjoint system (\ref{ade}).
\end{proof}

The equation $u_{n+2}-u_n=0$ and its adjoint equation $v_{n-2}-v_n=0$ are difference Euler-Lagrange equations with respect to a Lagrangian 
\begin{equation}
L_n=v_n(u_{n+2}-u_n).
\end{equation}

\begin{thm}\label{thmde}
Consider a system of difference equations (\ref{de}) and its adjoint system (\ref{ade}). If the original system (\ref{de}) admits a transformation group of symmetries with infinitesimal generator
\begin{equation}
X=Q^{\alpha}(n,[u])\frac{\partial}{\partial u^{\alpha}},
\end{equation}
then the combined system admits a transformation group of symmetries with an extended generator
\begin{equation}
Y=Q^{\alpha}(n,[u])\frac{\partial}{\partial u^{\alpha}}+Q_*^{\alpha}(n,[u],[v])\frac{\partial}{\partial v^{\alpha}}.
\end{equation} 
The functions $Q^{\alpha}_*(n,[u],[v])$ are to be determined (see (\ref{qx})).
\end{thm}
\begin{proof}
Write the Lagrangian as
\begin{equation}
L_n=v^{\beta}F_{\beta}(n,[u]),
\end{equation}
and consider the variational symmetry criterion, that is 
\begin{equation}\label{yyy}
\bold{pr}^{(\infty)}Y(L_n)=Y(v^{\beta})F_{\beta}+v^{\beta}\bold{pr}^{(\infty)}X(F_{\beta}).
\end{equation}
Recall that 
\begin{equation}
\bold{pr}^{(\infty)}X(F_{\beta})=\mathscr{B}_{\beta\alpha}(F_{\alpha}).
\end{equation}
The equation (\ref{yyy}) hence turns out to be
\begin{equation}
\begin{aligned}
\bold{pr}^{(\infty)}Y(L_n)=&Q^{\beta}_*F_{\beta}+\sum_{\alpha}v^{\beta}\mathscr{B}_{\beta\alpha}(F_{\alpha})\\
=&\sum_{\alpha}\left(Q_*^{\alpha}+\mathscr{B}_{\beta\alpha}^*(v^{\beta})\right)F_{\alpha}+\operatorname{Div}^{\vartriangle}R.
\end{aligned}
\end{equation}
Here $\mathscr{B}_{\beta\alpha}^*$, the adjoint operator of $\mathscr{B}_{\beta\alpha}$, and the tuple $R$ are obtained through the discrete version of integration by parts.
Therefore, the proof finishes by setting
\begin{equation}\label{qx}
Q_*^{\alpha}=-\mathscr{B}_{\beta\alpha}^*(v^{\beta}).
\end{equation}
\end{proof}

\begin{thm}
For the difference system (\ref{de}), each infinitesimal generator of symmetries $X=Q^{\alpha}\frac{\partial}{\partial u^{\alpha}}$ provides a conservation law for a system combined by the original system and its adjoint system.
\end{thm}
\begin{proof}
The proof is immediate. Theorem \ref{thmde} implies that a symmetry generator of the original system can be extended to a variational symmetry generator for a Lagrangian, which governs the combined system. Therefore, from Noether's theorem, we can get a conservation law, see (\ref{ncl}).
\end{proof}

\begin{exm}
Consider the following ordinary difference equation
\begin{equation}
u_{n+2}=\frac{u_nu_{n+1}}{2u_n-u_{n+1}}.
\end{equation}
It has been found that it admits a three-dimensional group of Lie point symmetries, whose characteristics are \cite{Hydon2000b} 
\begin{equation}
Q_1=u_n,~Q_2=nu_n^2,~Q_3=u_n^2.
\end{equation}
Define a Lagrangian
\begin{equation}
L_n=v_n\left(u_{n+2}-\frac{u_nu_{n+1}}{2u_n-u_{n+1}}\right).
\end{equation}
The adjoint equation is given by
\begin{equation}
v_n\frac{u_{n+1}^2}{(2u_n-u_{n+1})^2}-v_{n-1}\frac{2u_{n-1}^2}{(2u_{n-1}-u_n)^2}+v_{n-2}=0.
\end{equation}
For $Q_1$, we have that 
\begin{equation}
\begin{aligned}
&\left(Q_1\frac{\partial }{\partial u_n}+(SQ_1)\frac{\partial }{\partial u_{n+1}}+(S^2Q_1)\frac{\partial}{\partial u_{n+2}}\right)\left(u_{n+2}-\frac{u_nu_{n+1}}{2u_n-u_{n+1}}\right)\\
&~~~~~~~~~~~~~~~~~~~~~~~~~~~~~~~~~~~~~~~~~~~~~~~~~~~~~~~~~~~~~=u_{n+2}-\frac{u_nu_{n+1}}{2u_n-u_{n+1}}.
\end{aligned}
\end{equation}
From (\ref{qx}), this implies a group of variational symmetries for the Lagrangian, namely
\begin{equation}
Y_1=Q_1\frac{\partial}{\partial u_n}+Q_{1*}\frac{\partial}{\partial v_n} \text{ with } Q_{1*}=-v_n.
\end{equation}
This hence provides a conservation law (first integral) for a system combined by the original equation and its adjoint equation (see  (\ref{ncl})), namely
\begin{equation}
P_1=u_{n+1}v_{n-1}+u_nv_{n-2}-\frac{2u_nu_{n-1}^2}{(2u_{n-1}-u_n)^2}v_{n-1}.
\end{equation}
Here $R_1=0$, since
\begin{equation}
\bold{pr}^{(\infty)}Y_1(L_n)=0.
\end{equation}
Similarly, the other two extended characteristics with respect to the new coordinate $v_n$ are respectively
\begin{equation}
\begin{aligned}
&Q_{2*}=-(n+2)\left(u_{n+2}+\frac{u_nu_{n+1}}{2u_n-u_{n+1}}\right)v_n,\\
&Q_{3*}=-\left(u_{n+2}+\frac{u_nu_{n+1}}{2u_n-u_{n+1}}\right)v_n.
\end{aligned}
\end{equation}
We have $R_2=R_3=0$, and two first integrals
\begin{equation}
\begin{aligned}
&P_2=(n+1)u_{n+1}^2v_{n-1}+nu_n^2v_{n-2}-\frac{2nu_n^2u_{n-1}^2}{(2u_{n-1}-u_n)^2}v_{n-1},\\
&P_3=u_{n+1}^2v_{n-1}+u_n^2v_{n-2}-\frac{2u_n^2u_{n-1}^2}{(2u_{n-1}-u_n)^2}v_{n-1}.
\end{aligned}
\end{equation}
\end{exm}

\begin{exm}[{\bf The discrete KdV equation}]
A lattice version of the potential KdV equation
\begin{equation}
u_t=u_{xxx}+3u_x^2
\end{equation}
is given by a partial difference equation  (see for example \cite{Levi2007,Nijhoff1995})
\begin{equation}
(u_{0,0}-u_{1,1})(u_{1,0}-u_{0,1})+\beta-\alpha=0,
\end{equation}
which belongs to the ABS classification (equation H1) \cite{Adler2003}.
Here $u_{0,0}=u_{m,n}$ is the value of the dependent variable at the point $(m,n)\in \mathbb{Z}^2$, and $u_{i,j}$ denotes shifts of $u_{0,0}$. The arbitrary functions $\alpha$ and $\beta$ are assumed to be constants for simplicity. It admits a group of Lie point symmetries with the following infinitesimal generators \cite{Rasin2007b}
 \begin{equation}
\begin{aligned}
&X_1=\frac{\partial}{\partial u_{0,0}},~X_2=(-1)^{m+n}\frac{\partial}{\partial u_{0,0}},~X_3=(-1)^{m+n}u_{0,0}\frac{\partial}{\partial u_{0,0}},\\
&X_4=u_{0,0}\frac{\partial}{\partial u_{0,0}}+2\alpha\frac{\partial}{\partial \alpha}+2\beta\frac{\partial}{\partial \beta}.
\end{aligned}
\end{equation}
Let 
\begin{equation}
L_n=v_{0,0}\left[(u_{0,0}-u_{1,1})(u_{1,0}-u_{0,1})+\beta-\alpha\right],
\end{equation}
and we get the adjoint equation
\begin{equation}
(u_{1,0}-u_{0,1})v_{0,0}-(u_{0,-1}-u_{1,0})v_{0,-1}+(u_{-1,0}-u_{0,1})v_{-1,0}-(u_{0,-1}-u_{-1,0})v_{-1,-1}=0.
\end{equation}
From those generators, we get that
\begin{equation}
Q_{1*}=Q_{2*}=Q_{3*}=0,~Q_{4*}=-2v_n.
\end{equation}
It is not difficult to get that $R_1=R_2=R_3=R_4=0$ and the following conservation laws are obtained correspondingly 
\begin{equation}
\begin{aligned}
&\left\{\begin{array}{c}
P_1^1=S_{-1}\xi^1+S_{-1}\xi^3,
\vspace{0.2cm}\\
P_1^2=S_{-2}\xi^2+S_{-2}\xi^3,
\end{array}
\right.~
\left\{
\begin{array}{c}
P_2^1=(-1)^{m+n}(S_{-1}\xi^1-S_{-1}\xi^3),
\vspace{0.2cm}\\
P_2^2=(-1)^{m+n}(S_{-2}\xi^2-S_{-2}\xi^3),
\end{array}
\right.\\
&\left\{
\begin{array}{c}
P_3^1=(-1)^{m+n}(u_{0,0}S_{-1}\xi^1-u_{0,1}S_{-1}\xi^3),
\vspace{0.2cm}\\
P_3^2=(-1)^{m+n}(u_{0,0}S_{-2}\xi^2-u_{1,0}S_{-2}\xi^3),
\end{array}
\right.~
\left\{
\begin{array}{c}
P_4^1=u_{0,0}S_{-1}\xi^1+u_{0,1}S_{-1}\xi^3,
\vspace{0.2cm}\\
P_4^2=u_{0,0}S_{-2}\xi^2+u_{1,0}S_{-2}\xi^3.
\end{array}
\right.
\end{aligned}
\end{equation}
Here $S_{-i}$ denotes the first-order backward shift along the $i^{\text{th}}$-direction, and we write
\begin{equation}
\begin{aligned}
&\xi^1=(u_{0,0}-u_{1,1})v_{0,0}-(u_{1,-1}-u_{0,0})v_{0,-1},\\
&\xi^2=-(u_{0,0}-u_{1,1})v_{0,0}-(u_{0,0}-u_{-1,1})v_{-1,0},\\
&\xi^3=-(u_{1,0}-u_{0,1})v_{0,0}.
\end{aligned}
\end{equation}
Five-point symmetries are provided in \cite{Rasin2007b}, which will also lead to conservation laws. One may use even higher order symmetries.
\end{exm}

\section{Applications to ordinary/partial difference equations}
In this section, our method is applied to several ordinary and partial difference equations. By using known symmtries, we can get symmetries and conservation laws for the system combined by a difference system and its adjoint system. Conservation laws of the original system can be obtained if it satisfies a certain self-adjointness property. As well it is possible to construct conservation laws of the original system if special solutions of the adjoint system can be obtained.

\begin{exm}\label{exm41}
Consider a nonlinear difference equation 
\begin{equation}
u_{n+2}u_n-u_{n+1}^2=0,
\end{equation}
which admits a group of symmetries with characteristics \cite{Hydon2014} (see also \cite{Peng2014})
\begin{equation}
Q_1=u_n,~Q_2=nu_n.
\end{equation}
A Lagrangian can be defined as
\begin{equation}\label{exll}
L_n=v_n(u_{n+2}u_n-u_{n+1}^2)
\end{equation}
and hence we get its adjoint equation
\begin{equation}
v_nu_{n+2}-2v_{n-1}u_n+v_{n-2}u_{n-2}=0.
\end{equation}
The original difference equation is quasi self-adjoint, that can be verified by substituting $v_n=1/u_n^2$ in its adjoint equation.
For $Q_1$, we have that 
\begin{equation}
\left(Q_1\frac{\partial}{\partial u_n}+(SQ_1)\frac{\partial}{\partial u_{n+1}}+(S^2Q_1)\frac{\partial}{\partial u_{n+2}}\right)\left(u_{n+2}u_n-u_{n+1}^2\right)=2\left(u_{n+2}u_n-u_{n+1}^2\right).
\end{equation}
From (\ref{qx}), this implies an infinitesimal generator for the Lagrangian, namely
\begin{equation}
Y_1=Q_1\frac{\partial}{\partial u_n}+Q_{1*}\frac{\partial}{\partial v_n} \text{ with } Q_{1*}=-2v_n
\end{equation}
and hence a first integral is obtained via (\ref{ncl}) ($R_1=0$ here),
\begin{equation}
P_1=-2u_n^2v_{n-1}+u_{n+1}u_{n-1}v_{n-1}+u_nu_{n-2}v_{n-2}
\end{equation}
such that $SP_1=P_1$ on solutions of the Euler-Lagrange equations with respect to the Lagrangian (\ref{exll}). Letting $v_n=1/u_n^2$, the first integral becomes 
\begin{equation}
P_1=-2\frac{u_n^2}{u_{n-1}^2}+\frac{u_n}{u_{n-2}}+\frac{u_{n+1}}{u_{n-1}},
\end{equation}
which is unfortunately trivial, that is, $P_1=0$ on solutions of the original equation. 

Similarly, from $Q_2$, we get that
\begin{equation}
Q_{2*}=-2(n+1)v_n
\end{equation}
and
\begin{equation}
P_2=nu_n(-2u_nv_{n-1}+u_{n-2}v_{n-2})+(n+1)u_{n+1}u_{n-1}v_{n-1}.
\end{equation}
By setting $v_n=1/u_n^2$, the first integral becomes
\begin{equation}
P_2=-2n\frac{u_n^2}{u_{n-1}^2}+n\frac{u_n}{u_{n-2}}+(n+1)\frac{u_{n+1}}{u_{n-1}},
\end{equation}
which can be simplified into a nontrivial first integral
\begin{equation}\label{exm41fi}
P_2=\frac{u_n^2}{u_{n-1}^2}.
\end{equation}
\end{exm}

\begin{exm}\label{exlabc}
Consider a second-order linear ordinary difference equation
\begin{equation}
\left(-n+\frac{1}{2}\right)u_{n+2}+\left(2n-\frac{1}{2}\right)u_{n+1}-nu_n=0.
\end{equation}
A simple infinitesimal generator can be found that $X=u_n\frac{\partial}{\partial u_n}$. Introduce another variable $v_n$, and define a Lagrangian
\begin{equation}
L_n=v_n\left[\left(-n+\frac{1}{2}\right)u_{n+2}+\left(2n-\frac{1}{2}\right)u_{n+1}-nu_n\right].
\end{equation}
Its adjoint equation is hence
\begin{equation}
-nv_n+\left(2n-\frac{5}{2}\right)v_{n-1}+\left(-n+\frac{5}{2}\right)v_{n-2}=0.
\end{equation}
The generator $X$ is extended to a variational symmetry generator for the Lagrangian, namely $Y=X+Q_*\frac{\partial}{\partial v_n}$ with $Q_*=-v_n$. Therefore, a first integral via Noether's theorem is constructed (here $R=0$)
\begin{equation}
P=\left[\left(2n-\frac{5}{2}\right)v_{n-1}+\left(-n+\frac{5}{2}\right)v_{n-2}\right]u_n+\left(-n+\frac{3}{2}\right)v_{n-1}u_{n+1}.
\end{equation}
It is not difficult to see that the adjoint equation has a constant solution, that is $v_n=C$. Substituting this into the first integral $P$, we get a first integral of the original equation, that is,
\begin{equation}
nu_n+\left(-n+\frac{3}{2}\right)u_{n+1}.
\end{equation}
\end{exm}

The procedure in Example \ref{exlabc} can be applied to linear partial difference equations as well, since again their adjoint equations are independent from $[u]$.

\begin{exm}\label{exm43}
Consider the following linear partial difference equation
\begin{equation}\label{eqw}
\alpha(u_{1,0}+u_{-1,0})-\beta(u_{0,1}+u_{0,-1})=0,
\end{equation}
where $\alpha$ and $\beta$ are arbitrary positive constants.
It is a multisymplectic scheme for the nonlinear wave equation $u_{tt}-u_{xx}=0$ \cite{Bridges2001,Marsden1998,Peng2014b}. Provided that the constants $\alpha$ and $\beta$ are fixed, it admits a two-dimensional group of Lie point symmetries, whose infinitesimal generators are
\begin{equation}
X_1=u_{0,0}\frac{\partial}{\partial u_{0,0}},~X_2=(-1)^{m+n}u_{0,0}\frac{\partial}{\partial u_{0,0}}.
\end{equation}
Its adjoint equation is 
\begin{equation}
\alpha(v_{-1,0}+v_{1,0})-\beta(v_{0,-1}+v_{0,1})=0,
\end{equation}
and the governing Lagrangian is 
\begin{equation}
L_n=v_{0,0}\left[\alpha(u_{1,0}+u_{-1,0})-\beta(u_{0,1}+u_{0,-1})\right].
\end{equation}
It is obvious that the equation (\ref{eqw}) is self-adjoint. The infinitesimal generators are respectively extended to 
\begin{equation}
Y_1=X_1-v_{0,0}\frac{\partial}{\partial v_{0,0}},~Y_2=X_2+(-1)^{m+n}v_{0,0}\frac{\partial}{\partial v_{0,0}}.
\end{equation}
Since the Lagrangian depends on backward shifts of $u_{0,0}$, we may shift it forward and then apply the same procedure we used above. However, taking $Y_1$ as an example, we can reach a conservation law directly by
\begin{equation}
\begin{aligned}
\bold{pr}^{(\infty)}&Y_1(L_n)=-v_{0,0}\left[\alpha(u_{1,0}+u_{-1,0})-\beta(u_{0,1}+u_{0,-1})\right]\\
+&u_{0,0}\left[\alpha(v_{-1,0}+v_{1,0})-\beta(v_{0,-1}+v_{0,1})\right]\\
+&\alpha(S_1-\operatorname{id})(u_{0,0}v_{-1,0}-u_{-1,0}v_{0,0})+\beta(S_2-\operatorname{id})(u_{0,-1}v_{0,0}-u_{0,0}v_{0,-1}).
\end{aligned}
\end{equation}
Hence we get a conservation law for the combined system as
\begin{equation}
P_1^1=\alpha(u_{0,0}v_{-1,0}-u_{-1,0}v_{0,0}),~P_1^2=\beta(u_{0,-1}v_{0,0}-u_{0,0}v_{0,-1}).
\end{equation}
Setting $v=u$, this becomes a conservation law for the original equation. However, it is trivial. In particular, when $\alpha=\beta$, it is obvious that $v_{0,0}=1$ and $v_{0,0}=(-1)^{m+n}$ are two special solutions of the adjoint equation. Hence, respectively, we get two conservation laws for the original equation, namely
\begin{equation}
P_1^1=\alpha(u_{0,0}-u_{-1,0}),~P_1^2=\alpha(u_{0,-1}-u_{0,0})
\end{equation}
and
\begin{equation}
P_1^1=\alpha(-1)^{m+n+1}(u_{0,0}+u_{-1,0}),~P_1^2=\alpha(-1)^{m+n}(u_{0,-1}+u_{0,0}).
\end{equation}

Similarly, for $Y_2$, we can get a conservation law for the combined system
\begin{equation}
\left\{\begin{array}{c}
P_2^1=\alpha(-1)^{m+n}(u_{0,0}v_{-1,0}+u_{-1,0}v_{0,0}),
\vspace{0.2cm}\\
P_2^2=\beta(-1)^{m+n+1}(u_{0,-1}v_{0,0}+u_{0,0}v_{0,-1}).
\end{array}
\right.
\end{equation}
Let $u=v$, and it becomes a conservation law for the original system, namely
\begin{equation}\label{exm43cl}
P_2^1=2\alpha(-1)^{m+n}u_{0,0}u_{-1,0},~
P_2^2=2\beta(-1)^{m+n+1}u_{0,-1}u_{0,0}.
\end{equation}
When $\alpha=\beta$, again we can follow the same procedure as that for $Y_1$ to obtain conservation laws. However, those conservation laws via substitutions $v_{0,0}=1$ and $v_{0,0}=(-1)^{m+n}$ are equivalent to the ones we already obatined. 
\end{exm}

By applying symmetries to known conservation laws, it is possible to construct new conservation laws, which is called a symmetry method for conservation laws in \cite{Rasin2009}. Let $X$ be an infinitesimal generator and $P$ be a conservation law for some difference system. Since the prolonged symmetry generator and shift operators commute with each other, i.e.
\begin{equation}
[\bold{pr}^{(\infty)}X,S_J]=0, \text{ for all $J$},
\end{equation}
we have that $\bold{pr}^{(\infty)}X(P)$ is again a conservation law. Therefore, we can use one symmetry generator again and again to obtain more conservation laws. In particular, if a difference system admits infinitely many symmetries, then we may construct infinityly many conservation laws either by using the method developed in this paper or the symmetry method. Though no promise is given that such new conservation laws are independent from one another. Neither are they nontrivial. For example, the two symmetry generators in Example \ref{exm41} and the first integral (\ref{exm41fi}) will lead to either trivial or equivalent first integrals. Similar consequence happens for Example \ref{exlabc} and Example \ref{exm43}. Nevertheless, higher order symmetries are usually helpful.

\begin{exm}\label{exm44}
Let us consider the equation in Example \ref{exm43} again. It admits infinitely many symmtries with infinitesimal generators
\begin{equation}
X_{ij}=u_{i,j}\frac{\partial}{\partial u_{0,0}}
\end{equation}
and
\begin{equation}
\widetilde{X}_{ij}=(-1)^{m+n}u_{i,j}\frac{\partial }{\partial u_{0,0}},
\end{equation}
for any integers $i$ and $j$. In this situation, both our new method and the symmetry method will lead to infinitely many conservation laws. Here since we already obtained a conservation law via our method  in Example \ref{exm43}, the symmetry method seems more immediate. Applying $\bold{pr}^{(\infty)}X_{ij}$ to the conservation law (\ref{exm43cl}), we obtain infinitely many conservation laws
\begin{equation}
\left\{\begin{array}{c}
P_{ij}^1=2\alpha(-1)^{m+n}(u_{-1,0}u_{i,j}+u_{0,0}u_{i-1,j}),
\vspace{0.2cm}\\
P_{ij}^2=2\beta(-1)^{m+n+1}(u_{0,-1}u_{i,j}+u_{0,0}u_{i,j-1}).
\end{array}\right.
\end{equation}
By using $X_{ij}$ and the new conservation laws $P_{kl}$, we can get even more conservation laws, namely
\begin{equation}
\left\{\begin{array}{c}
P_{ijkl}^1=2\alpha(-1)^{m+n}(u_{i-1,j}u_{k,l}+u_{i,j}u_{k-1,l}),
\vspace{0.2cm}\\
P_{ijkl}^2=2\beta(-1)^{m+n+1}(u_{i,j-1}u_{k,l}+u_{i,j}u_{k,l-1}).
\end{array}\right.
\end{equation}
We may continue this procedure, though it is possible that the newly obtained conservation laws will be trivial or equivalent to the ones we already got. Similarly for $\widetilde{X}_{ij}$, we get
\begin{equation}
\left\{\begin{array}{c}
\widetilde{P}_{ij}^1=2\alpha(u_{-1,0}u_{i,j}-u_{0,0}u_{i-1,j}),
\vspace{0.2cm}\\
\widetilde{P}_{ij}^2=2\beta(u_{0,0}u_{i,j-1}-u_{0,-1}u_{i,j}),
\end{array}\right.
\end{equation}
and
\begin{equation}
\left\{\begin{array}{c}
\widetilde{P}_{ijkl}^1=2\alpha(-1)^{m+n+k+l}(u_{-1,0}u_{i+k,j+l}+u_{0,0}u_{i+k-1,j+l}),
\vspace{0.2cm}\\
\widetilde{P}_{ijkl}^2=2\beta(-1)^{m+n+k+l}(u_{0,-1}u_{i+k,j+l}+u_{0,0}u_{i+k,j+l-1}).
\end{array}\right.
\end{equation}
For each pair of fixed $k$ and $l$, we can always find $\widetilde{P}_{ijkl}$ from the set $\{P_{ij}\}$ (with a multiplication of $(-1)^{k+l}$), that is, the expression of $\widetilde{P}_{ijkl}$ provides no more nonequivalent conservation laws.
\end{exm}

\section*{ Acknowledgements} This work is supported by JST-CREST.

\vspace{0.3cm} 
L.Peng@aoni.waseda.jp

Department of Applied Mechanics and Aerospace
Engineering, Waseda University, Okubo, Shinjuku, Tokyo 169-8555,
Japan

\end{document}